\documentclass[5p,times,twocolumn]{elsarticle}

\usepackage{amsmath,amssymb,amsfonts,color}
\usepackage{graphicx}
\usepackage{textcomp}
\usepackage{xcolor}
\usepackage{enumerate}
\usepackage{setspace}
\usepackage{bm}
\usepackage{comment}
\usepackage{caption}
\usepackage{tipa}
\usepackage{footnote}
\usepackage{subfigure}
\usepackage{algorithm}
\usepackage{algorithmic}
\usepackage{url}
\usepackage{multirow}

\def\x{{\mathbf x}}

\usepackage{xcolor}
\begin{document}

\begin{frontmatter}

\title{Boosting Automatic COVID-19 Detection Performance with Self-Supervised Learning and Batch Knowledge Ensembling}

\author{Guang Li${}^\text{a}$}
\ead{guang@lmd.ist.hokudai.ac.jp}
\author{Ren Togo${}^\text{b}$}
\ead{togo@lmd.ist.hokudai.ac.jp}
\author{Takahiro Ogawa${}^\text{b}$}
\ead{ogawa@lmd.ist.hokudai.ac.jp}
\author{Miki Haseyama${}^\text{b}$}
\ead{mhaseyama@lmd.ist.hokudai.ac.jp}
\address{${}^\text{a}$Graduate School of Information Science and Technology, Hokkaido University, \\
           N-14, W-9, Kita-Ku, Sapporo, 060-0814, Japan}
\address{${}^\text{b}$Faculty of Information Science and Technology, Hokkaido University, \\
           N-14, W-9, Kita-Ku, Sapporo, 060-0814, Japan}

\begin{abstract}

{\it Problem:~}
\textcolor{black}{
Detecting COVID-19 from chest X-Ray (CXR) images has become one of the fastest and easiest methods for detecting COVID-19.
However, the existing methods usually use supervised transfer learning from natural images as a pretraining process. 
These methods do not consider the unique features of COVID-19 and the similar features between COVID-19 and other pneumonia.
}
{\it Aim:~}
\textcolor{black}{
In this paper, we want to design a novel high-accuracy COVID-19 detection method that uses CXR images, which can consider the unique features of COVID-19 and the similar features between COVID-19 and other pneumonia.
}

{\it Methods:~}
\textcolor{black}{
Our method consists of two phases. 
One is self-supervised learning-based pertaining; the other is batch knowledge ensembling-based fine-tuning.
Self-supervised learning-based pretraining can learn distinguished representations from CXR images without manually annotated labels.
On the other hand, batch knowledge ensembling-based fine-tuning can utilize category knowledge of images in a batch according to their visual feature similarities to improve detection performance.
Unlike our previous implementation, we introduce batch knowledge ensembling into the fine-tuning phase, reducing the memory used in self-supervised learning and improving COVID-19 detection accuracy.
}

{\it Results:~}
\textcolor{black}{
On two public COVID-19 CXR datasets, namely, a large dataset and an unbalanced dataset, our method exhibited promising COVID-19 detection performance.
Our method maintains high detection accuracy even when annotated CXR training images are reduced significantly (e.g., using only 10\% of the original dataset).
In addition, our method is insensitive to changes in hyperparameters.
}

{\it Conclusion:~} 
\textcolor{black}{
The proposed method outperforms other state-of-the-art COVID-19 detection methods in different settings.
Our method can reduce the workloads of healthcare providers and radiologists.
}

\end{abstract}


\end{frontmatter}

\section{Introduction}
\label{sec1}
As a pandemic, COVID-19 has rapidly spread worldwide, affecting the health and lives of billions of people~\cite{andersen2020proximal}.
There have been 632,533,408 confirmed cases of COVID-19, including 6,592,320 deaths reported in the world as of November 15, 2022.\footnote{https://covid19.who.int}
\textcolor{black}{
Vaccines have played a critical role in preventing the spread of new infections, but the situation remains critical because some countries and special populations are still unable to receive the vaccine.
Recently, the number of infected people has been increasing drastically since the high variability of COVID-19.
}
Identifying infected patients early and separating them from the population is crucial to controlling the infection.
In line with WHO reports, the gold standard for COVID-19 detection is currently real-time reverse transcription polymerase chain reaction (RT-PCR)~\cite{pujadas2020comparison}.
Although RT-PCR has many advantages, it also has some shortcomings, including a high false-negative rate and being a lengthy process~\cite{drame2020should}.
Furthermore, RT-PCR is typically inadequate in many hard-hit and underdeveloped areas, which is detrimental to stop the continued spread of COVID-19 worldwide~\cite{li2020role}.
The use of computed tomography (CT) and X-ray have been complementary to RT-PCR for COVID-19 detection~\cite{rubin2020role}.
\par
The common presence of radiologic findings of pneumonia in patients with COVID-19 makes radiologic examinations valuable for COVID-19 detection~\cite{shi2020radiological}.
The chest X-ray (CXR) is particularly beneficial because it cost less, takes less time, and exposes you to less radiation than CT~\cite{chen2022classification}.
The detection of COVID-19 through chest radiography offers significant potential for screening and analyzing the health of patients.
However, manual detection of COVID-19 from CXR images has several problems.
For example, requiring radiologists to detect COVID-19 infection from several CXR images is challenging as it is time-consuming and easily has human errors~\cite{shoeibi2020automated}.
Furthermore, it is difficult to distinguish COVID-19 from other viral pneumonia cases because of their similar appearances~\cite{chaddad2021deep}.
\par
To overcome the above problems, several researchers have used deep learning (DL) to detect COVID-19 infection from CXR images~\cite{kumar2022novel, gulakala2022rapid}.
A study~\cite{minaee2020deep} has proposed using transfer learning with dominant convolutional neural networks (CNNs) such as ResNet~\cite{he2016deep}, SqueezeNet~\cite{iandola2017squeezenet}, and DenseNet~\cite{huang2017densely} to perform COVID-19 detection from CXR images, achieving good detection performance on an unbalanced small COVID-19 dataset.
Moreover, another study~\cite{ismael2021deep} has proposed the ensemble fine-tuning of CNNs, end-to-end training of CNNs, and deep feature extraction for an additional support vector machine (SVM) classifier to obtain better COVID-19 detection performance.
\textcolor{black}{
However, these methods used supervised transfer learning from natural images as a pretraining process and did not consider the unique features of COVID-19 and the similar features between COVID-19 and other pneumonia.
Furthermore, these studies typically evaluated small COVID-19 CXR image datasets, which may have limitations when used in a real-world clinical situation.
}
\par
\textcolor{black}{
In this study, we propose a novel automatic COVID-19 detection method with self-supervised learning and batch knowledge ensembling using CXR images.
Self-supervised learning can learn distinguished representations from CXR images without manually annotated labels as a pretraining phase, which is more suitable for complex COVID-19 visual features than supervised learning.
Furthermore, batch knowledge ensembling can utilize category information of images in a batch according to their visual feature similarities to boost detection performance.
}
The proposed method achieved promising COVID-19 detection performance on a large COVID-19 CXR dataset and an unbalanced COVID-19 CXR dataset.
Our method also maintains high detection accuracy even when the annotated CXR training images are reduced significantly (e.g., using only 10\% of the original dataset).
In addition, our method is insensitive to changes in hyperparameters.
As a result, the proposed method outperforms state-of-the-art (SOTA) COVID-19 detection methods in different settings.
Our method can reduce the workloads of healthcare providers and radiologists.
\par
Our contributions are summarized as follows.
\begin{itemize}
    \item We propose performing self-supervised learning to learn distinguished representations without manually annotated labels from CXR images as a pretraining phase, which is more suitable for complex COVID-19 visual features than supervised learning.
    \item We propose batch knowledge ensembling-based fine-tuning, which can utilize category information of images in a batch according to their visual feature similarities to boost the detection performance.
    \item The proposed method achieves high detection accuracy on a large COVID-19 CXR dataset and an unbalanced dataset; it is insensitive to changes in hyperparameters.
    \item The proposed method can maintain high detection accuracy even when the annotated CXR training images are reduced significantly.
\end{itemize}
\par
A preliminary report of this study was published in our previous study~\cite{li2022self}.
Our previous work is extended in this paper in the following ways. 
First, we introduce batch knowledge ensembling into the fine-tuning phase, reducing the memory used in self-supervised learning and achieving higher COVID-19 detection accuracy.
Second, we evaluate the performance of the proposed method in two network structures compared with other SOTA methods.
Third, we use a new unbalanced COVID-19 dataset to evaluate the robustness of the proposed method.
Finally, we discuss the effects of hyperparameters on the proposed method.
\par
\textcolor{black}{
The remainder of this paper is organized as follows. 
In Section~\ref{sec2}, we introduce related works.
In Section~\ref{sec3}, we describe the details of the proposed method.
Then, we present the experiments, discussion, and conclusions in Sections~\ref{sec4},~\ref{sec5}, and~\ref{sec6}, respectively.
}
\section{Related Works}
\label{sec2}
\subsection{Automatic COVID-19 Detection}
\label{2.1}
\textcolor{black}{
Numerous studies have been conducted with DL to perform COVID-19 detection using CXR~\cite{subramanian2022review} or CT images~\cite{mohammed2022novel}. 
Different neural network structures, transfer learning techniques, and ensemble methods have been proposed to improve the performance of automatic COVID-19 detection.
For example, several studies~\cite{rahaman2020identification, taresh2021transfer, jin2021hybrid, ibrahim2021pneumonia, umar2022convolutional, nagi2022performance} have proposed transfer learning methods with the most used deep CNNs such as AlexNet~\cite{krizhevsky2012imagenet},  VGGNet~\cite{simonyan2014very}, InceptionNet~\cite{szegedy2016rethinking}, ResNet, DenseNet, CheXNet~\cite{rajpurkar2017chexnet}, and MobileNet~\cite{tan2019mnasnet} to perform COVID-19 detection from CXR images.
Furthermore, some researchers~\cite{hasoon2021covid, mohammed2021comprehensive, gayathri2022computer, aslan2022covid, mohammed2021intelligent} have proposed integrating DCNN feature extraction with machine learning algorithms, such as SVM~\cite{ben2010user}, k-nearest neighbour~\cite{gou2019generalized}, naive bayes~\cite{webb2010naive}, decision tree~\cite{myles2004introduction}, reliefF~\cite{robnik2003theoretical}, and sparse autoencoder~\cite{xu2015stacked}.
In addition, several studies have used capsule networks or attention-based networks to perform COVID-19 detection from CXR images, which can consider more spatial relationships than traditional DCNNs~\cite{tiwari2021convolutional, afshar2020covid, toraman2020convolutional, lin2021aanet}.
Furthermore, some studies have proposed using confidence-aware or uncertainty-aware learning for robust COVID-19 detection~\cite{zhang2020viral, shamsi2021uncertainty, dong2021rconet}.
Although these methods achieved good COVID-19 detection performance, they were typically tested on small COVID-19 datasets and applied only to supervised learning.
}
\begin{figure*}[t]
        \centering
        \includegraphics[width=16cm]{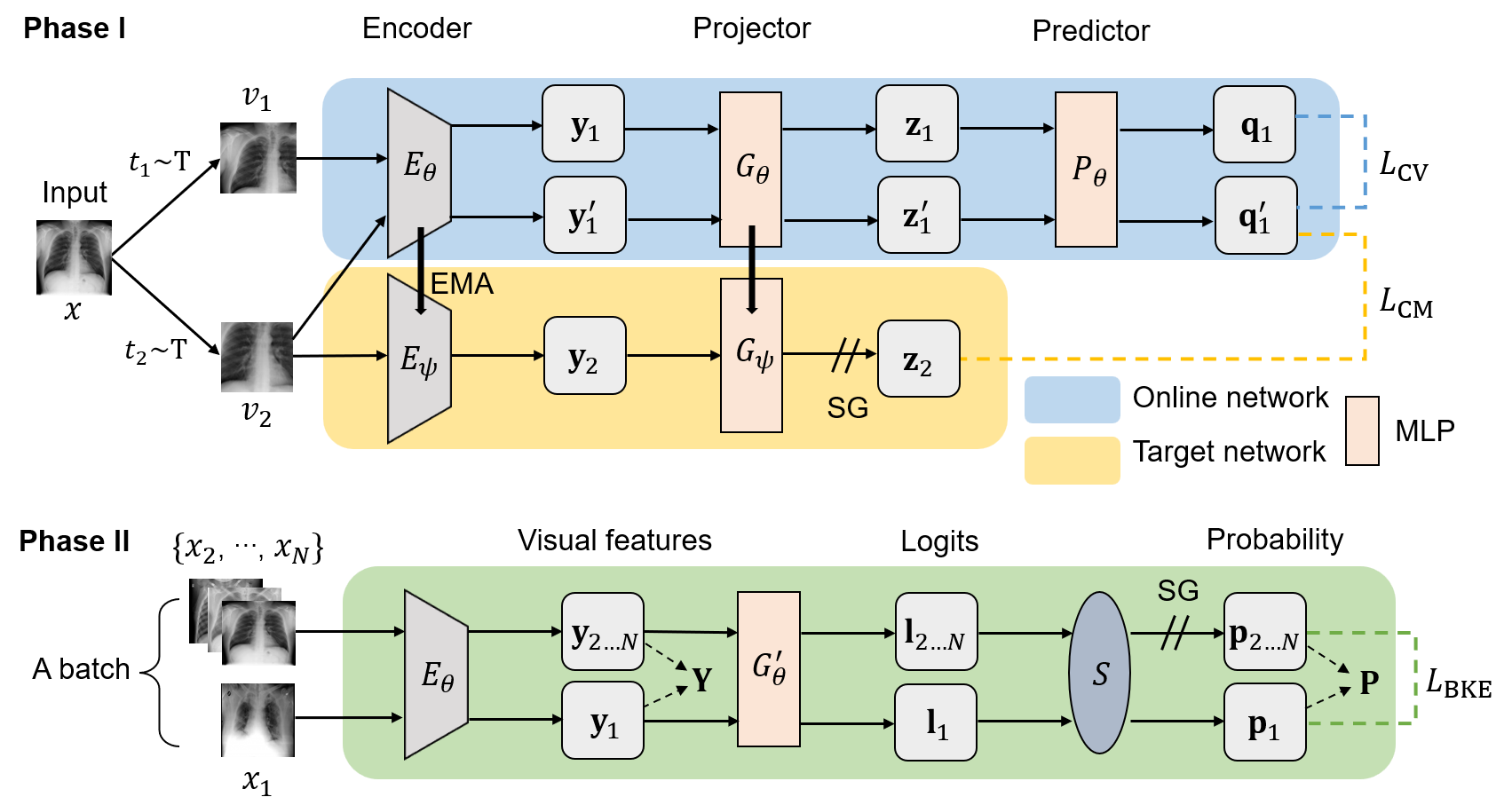}
        \caption{Overview of the proposed method. Our method consists of two phases, Phase I is self-supervised learning-based pretraining, and Phase II is batch knowledge ensembling-based fine-tuning. $\mathrm{EMA}$ represents exponential moving average, $\mathrm{SG}$ represents stop-gradient, $\mathrm{MLP}$ represents multilayer perceptron, and $S$ represents softmax function.}
        \label{fig1}
\end{figure*}
\subsection{Self-Supervised Learning}
\label{2.2}
\textcolor{black}{
In the past few years, self-supervised learning has attracted widespread attention in the field of machine learning~\cite{liu2021self}.
In contrast to supervised learning, self-supervised learning takes advantage of image characteristics (such as position, color, and texture) without manually annotating labels~\cite{minderer2020automatic}.
Studies have demonstrated that good representations can be learned from predicting context or playing a jigsaw game on images~\cite{doersch2015unsupervised, noroozi2016unsupervised}.
Furthermore, some contrastive-based self-supervised learning methods are effective on natural image datasets~\cite{chen2020simple, grill2020bootstrap, chen2021exploring}.
A Siamese network is used to maximize the similarity between the representations of views, with inputs defined as two augmented views from one image~\cite{tian2020understanding, pantazis2021focus}.
Compared with supervised learning, self-supervised learning without manually labeled annotations can learn fine-grained representations and is suitable for high-complexity CXR images~\cite{zhou2019models, azizi2021big}.
Therefore, we propose learning distinguished representations from COVID-19 CXR images using contrastive-based self-supervised learning as a pretraining phase. 
}
\subsection{Knowledge Ensembling}
\label{2.3}
\textcolor{black}{
Generally, an ensemble of multiple networks produces better predictions than a single network~\cite{hinton2015distilling}.
By aggregating different networks, knowledge ensembling technologies generate robust supervision signals~\cite{gou2021knowledge}.
Knowledge ensembling even has been applied to semi-supervised~\cite{tarvainen2017mean} and self-supervised scenarios~\cite{grill2020bootstrap, he2020momentum}.
However, the existing knowledge ensembling is limited to outputs of multi-networks, which is not applicable in low computation resource situations such as clinical use.
Hence, we proposed a novel batch knowledge ensembling method, which utilizes the knowledge in the data of a training batch with only one network.
With batch knowledge ensembling-based fine-tuning, our method can boost COVID-19 detection accuracy with a low-computing source.
}
\section{COVID-19 Detection with Self-Supervised Learning and Batch Knowledge Ensembling}
\label{sec3}
Figure~\ref{fig1} shows an overview of the proposed method.
\textcolor{black}{
Our method consists of two phases: the first is a self-supervised learning-based fine-tuning phase for learning distinguished representations from CXR images, and the second is a batch knowledge ensembling-based fine-tuning phase for the accurate automatic detection of COVID-19.
}
We show the details of the first and second phases in subsections~\ref{3.1} and~\ref{3.2}, respectively.
\subsection{Phase I: Self-Supervised Learning-based Pretraining}
\label{3.1}
First, we introduce the self-supervised learning-based pretraining phase.
\textcolor{black}{
Our self-supervised learning method uses an online network and a target network to learn distinguished representations from CXR images.
Encoder $E_{\theta}$, projector $G_{\theta}$, and predictor $P_{\theta}$ belong to the online network.
Encoder $E_{\psi}$ and projector $G_{\psi}$ belong to the target network.
The transformations $t_{1}$ and $t_{2}$ for an input CXR image $x$ are chosen at random from the distribution $T$ to obtain a pair of views $v_{1} = t_{1}(x_{1})$ and $v_{2} = t_{2}(x_{2})$.
Various augmentation methods are used in these transformations, including cropping, resizing, flipping, color jittering, and Gaussian blur.
}
\par
Online network encoder $E_{\theta}$ and projector $G_{\theta}$ process the view $v_{1}$.
Accordingly, target network encoder $E_{\psi}$ and projector $G_{\psi}$ process the view $v_{2}$.
As a result, the output $\mathbf{z}_{2}$ represents the feature of $v_{2}$ in the target network.
For cross-view loss calculations, we use $v_{2}$ to obtain a copy and input it into the online network.
Subsequently, we transform two views into features $\mathbf{q}_{1}$ and $\mathbf{q}'_{1}$ using the predictor $P_{\theta}$ of the online network.
The cross-view loss $L_{\mathrm{CV}}$ can be calculated as follows:
\begin{equation}
\begin{split}
L_{\mathrm{CV}} 
& = || \hat{\mathbf{q}}_{1} - \hat{\mathbf{q}}'_{1} ||_{2}^{2}
\\ & = 2 - 2 \cdot \frac{\left \langle \mathbf{q}_{1},\mathbf{q}'_{1} \right \rangle}{ || \mathbf{q}_{1} ||_{2} \cdot || \mathbf{q}'_{1} ||_{2}},
\end{split}
\end{equation}
where $\hat{\mathbf{q}}_{1} = \mathbf{q}_{1}/ || \mathbf{q}_{1} ||_{2}$ and $\hat{\mathbf{q}}'_{1} = \mathbf{q}'_{1}/ || \mathbf{q}'_{1} ||_{2}$ represent the normalized features of $v_{1}$ and $v_{2}$ processed by the online network, respectively.
The cross-model loss $L_{\mathrm{CM}}$ can be calculated as follows:
\begin{equation}
\begin{split}
L_{\mathrm{CM}} 
& = || \hat{\mathbf{q}}'_{1} - \hat{\mathbf{z}}_{2} ||_{2}^{2} 
\\ & = 2 - 2 \cdot \frac{\left \langle \mathbf{q}'_{1}, \mathbf{z}_{2} \right \rangle}{ || \mathbf{q}'_{1} ||_{2} \cdot || \mathbf{z}_{2} ||_{2}},
\end{split}
\end{equation}
where $\hat{\mathbf{z}}_{2} = \mathbf{z}_{2}/ || \mathbf{z}_{2} ||_{2}$ represents the normalized features of $v_{2}$ processed by the target network.
The predictor $P_{\theta}$ is used only in the online network to prevent learning collapse~\cite{grill2020bootstrap}.
By minimizing the total loss, the weights of the online network ($\theta$) are updated.
The total loss $\mathcal{L}_{\theta,\psi}$ and optimizing process are defined as follows:
\begin{equation}
\label{equ3}
\mathcal{L}_{\theta,\psi} = \mathcal{L}_{\mathrm{CV}} + \mathcal{L}_{\mathrm{CM}},
\end{equation}
\begin{equation}
\theta \leftarrow \mathrm{Opt}(\theta, \nabla_{\theta}\mathcal{L}_{\theta,\psi}, \alpha),
\end{equation}
where $\mathrm{Opt}$ represents the optimizer and $\alpha$ represents the learning rate.
Based on the weights of the online network ($\theta$), the target network weights ($\psi$) are derived by multiplying them by the exponential moving average.
\begin{equation}
\psi \leftarrow \zeta\psi + (1-\zeta)\theta,
\end{equation}
where $\zeta$ represents the degree of moving average.
The gradient is not backpropagated through the target network for stable training~\cite{chen2021exploring}.
\par
\textcolor{black}{
After the self-supervised learning-based pretraining phase, the online network encoder $E_{\theta}$ learns distinguished representations from CXR images, and the parameters are saved for the next fine-tuning phase.
Compared with supervised learning, self-supervised learning without manually labeled annotations can learn fine-grained representations and is suitable for high-complexity CXR images.
}
\subsection{Phase II: Batch Knowledge Ensembling-based Fine-tuning}
\label{3.2}
Next, we introduce the batch knowledge ensembling-based fine-tuning phase of our method.
\textcolor{black}{
Combined knowledge from images with similar visual features could provide better classification performance because they tend to have similar predicted probabilities.   
The above concept can be used to improve COVID-19 detection performance according to the similarity of visual features between different CXR images in a batch.
}
\par
First, we can obtain the visual feature similarity matrix $\mathbf{Y} \in \mathbb{R}^{N \times N}$ using the encoded visual features $\{\mathbf{y}_{1}, ..., \mathbf{y}_{N}\}$ in a batch ($N$ images) as follows:
\begin{equation}
\mathbf{Y}_{i,j} = (\hat{\mathbf{y}}_{i}^{\top} \hat{\mathbf{y}}_{j}),
\end{equation}
where $\hat{\mathbf{y}}_{i} = \mathbf{y}_{i}/ || \mathbf{y}_{i} ||_{2}$ represents the normalized features, and $i,j$ represents the batch indices.
To avoid self-knowledge reinforcement, we eliminate diagonal entries with the identity matrix $\mathbf{I}$ by $ \mathbf{Y} = \mathbf{Y} \odot (1-\mathbf{I})$.
Then, we normalize the similarity matrix for visual features as follows:
\begin{equation}
\hat{\mathbf{Y}}_{i,j} = \frac{\mathrm{exp}(\mathbf{Y}_{i,j})}{\sum_{j \neq i}\mathrm{exp}(\mathbf{Y}_{i, j})}, \forall i \in \{1,...,N\}.
\end{equation}
By applying a projector $G'_{\theta}$ and a softmax function $S$ to the output logits $\{\mathbf{l}_{1}, ..., \mathbf{l}_{N}\}$, we can obtain the predictive probabilities $\{\mathbf{p}_{1}, ..., \mathbf{p}_{N}\}$ as follows:
\begin{equation}
\mathbf{p}_{(k)} = \frac{\mathrm{exp}(\mathbf{l}_{k}/\tau)}{\sum_{i = 1}^{K}\mathrm{exp}(\mathbf{l}_{i}/\tau)},
\end{equation}
where $K$ represents the number of classes, and $\tau$ represents a temperature hyperparameter for the soft degree. 
The probability matrix of a batch of CXR images is predicted as $\mathbf{P} = [\mathbf{p}_{1}, ..., \mathbf{p}_{N}]^{\top} \in \mathbb{R}^{N \times K}$.
We generate the soft targets $\mathbf{Q}$ as a weighted sum of the initial probability matrix $\mathbf{P}$ and the propagated probability matrix $\hat{\mathbf{Y}}\mathbf{P}$ to prevent propagating noisy predictions as follows:
\begin{equation}
\mathbf{Q} = \omega\hat{\mathbf{Y}}\mathbf{P} + (1-\omega)\mathbf{P}.
\end{equation}
Furthermore, we propagate and ensemble multiple times to generate better soft targets $\mathbf{Q}$ for batch knowledge ensembling as follows:
\begin{equation}
\begin{split}
\mathbf{Q}_{(t)}
& = \omega\hat{\mathbf{Y}}\mathbf{Q}_{(t-1)} + (1-\omega)\mathbf{P},
\\ & = (\omega\hat{\mathbf{Y}})^{t}\mathbf{P} + (1-\omega)\sum_{i = 0}^{t-1}(\omega\hat{\mathbf{Y}})^{i}\mathbf{P},
\end{split}
\end{equation}
where $\omega$ represents a weight factor, and $t$ represents the $t$-th iteration.
As the number of iterations approaches infinite, we obtain $\mathrm{lim}_{t\rightarrow\infty}(\omega\hat{\mathbf{Y}})^{t} = 0$ and $\mathrm{lim}_{t\rightarrow\infty}\sum_{i = 0}^{t-1}(\omega\hat{\mathbf{Y}})^{i} = (\mathbf{I} - \omega\hat{\mathbf{Y}})^{-1}$.
Based on the above observation, an approximate inference formulation can be calculated as follows:
\begin{equation}
\mathbf{Q} = (1-\omega)(\mathbf{I} - \omega\hat{\mathbf{Y}})^{-1}\mathbf{P}.
\end{equation}
Finally, we define the batch knowledge ensembling loss $L_{\mathrm{BKE}}$ as follows:
\begin{equation}
L_{\mathrm{BKE}} = L_{\mathrm{CE}} + \lambda \cdot \tau^{2} \cdot D_\mathrm{KL}(\mathbf{Q}||\mathbf{P}),
\end{equation}
where $L_{\mathrm{CE}}$ represents the ordinary cross-entropy loss, $D_\mathrm{KL}$ represents Kullback-Leibler divergence, and $\lambda$ represents a balance hyperparameter.
There is no backpropagation of gradients through soft targets for stable training~\cite{li2022self}.
\par
\textcolor{black}{
Our previous method~\cite{li2022self} introduces batch knowledge ensembling into the self-supervised learning phase, making it memory costly and sensitive to hyperparameters.
Unlike the previous implementation, we introduce batch knowledge ensembling into the fine-tuning phase, reducing the memory used in self-supervised learning and improving COVID-19 detection accuracy.
According to the similarities of visual features in the different CXR images, the encoder $E_{\theta}$ can learn better representations and use them for the final COVID-19 detection.
}
\section{Experiments}
\label{sec4}
\subsection{Dataset and Settings}
\begin{figure}[t]
        \centering
        \subfigure[]{
        \centering
        \includegraphics[width=4.0cm]{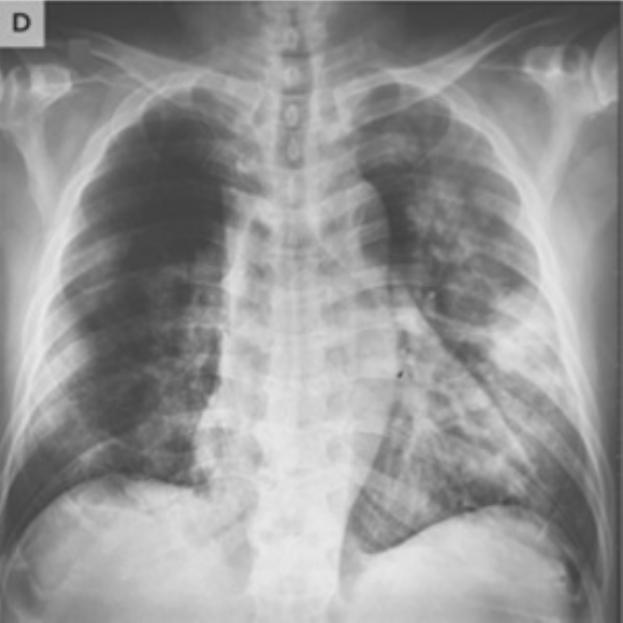}
        }
        \subfigure[]{
        \centering
        \includegraphics[width=4.0cm]{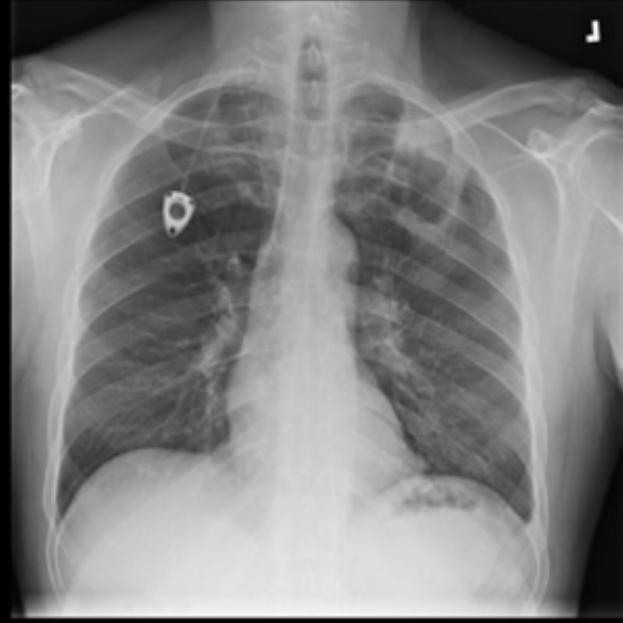}
        }
        \subfigure[]{
        \centering
        \includegraphics[width=4.0cm]{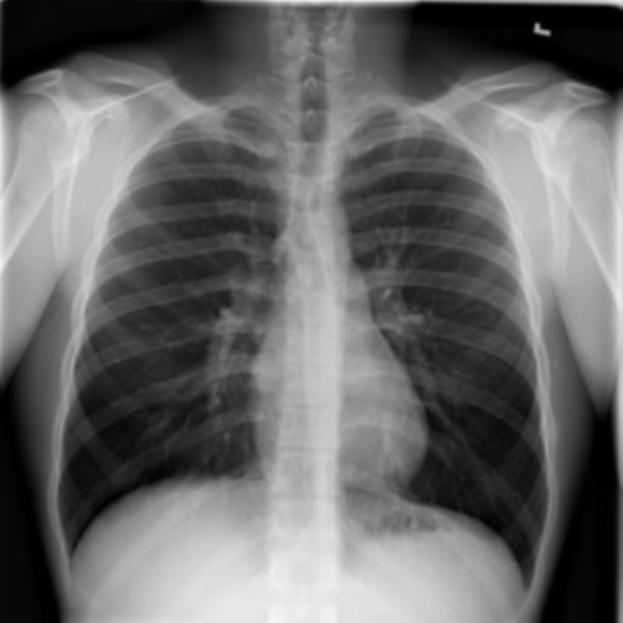}
        }
        \subfigure[]{
        \centering
        \includegraphics[width=4.0cm]{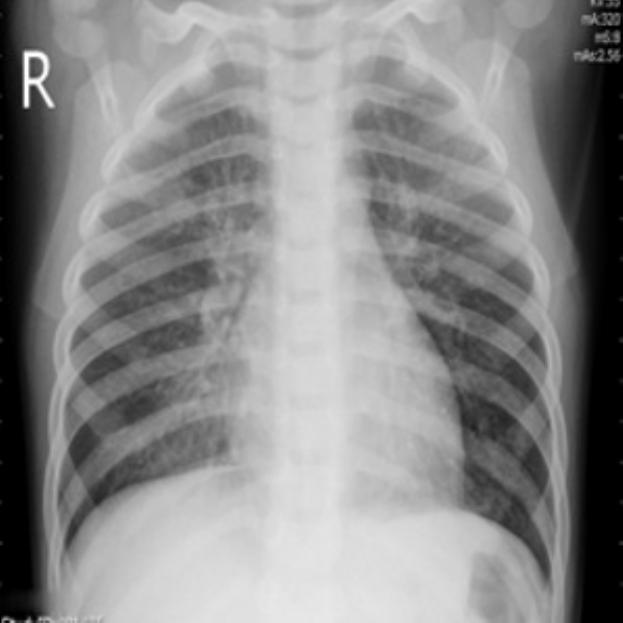}
        }
        \caption{Examples of CXR images in the large COVID-19 CXR dataset~\cite{rahman2021exploring}: (a) COVID-19, (b) Lung Opacity (c) Normal, and (d) Viral Pneumonia.}
        \label{fig2}
\end{figure}
\begin{figure}[t]
        \centering
        \subfigure[]{
        \centering
        \includegraphics[width=4.0cm]{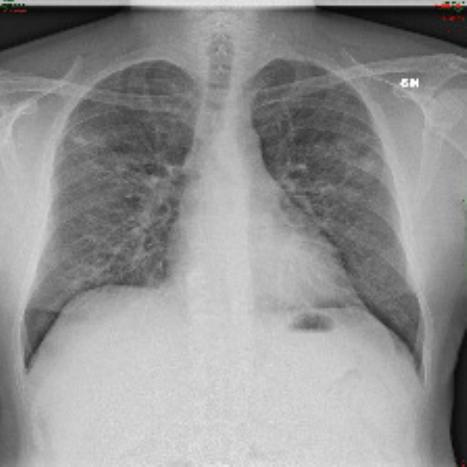}
        }
        \subfigure[]{
        \centering
        \includegraphics[width=4.0cm]{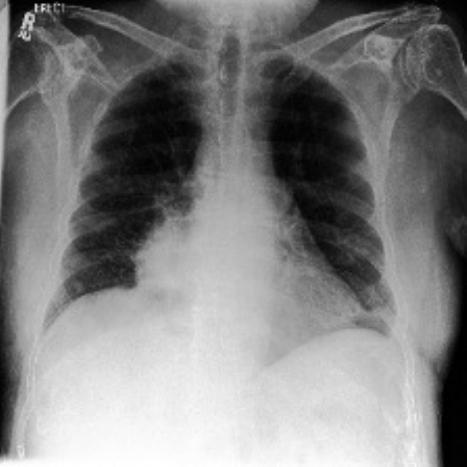}
        }
        \caption{Examples of CXR images in the COVID5K dataset~\cite{minaee2020deep}: (a) COVID-19 , (b) Normal.}
        \label{fig3}
\end{figure}
\begin{table}[t]
    \centering
    \caption{Details of the large COVID-19 CXR dataset~\cite{rahman2021exploring}.}
    \label{tab1}
    \begin{tabular}{lccc}
    \hline
    Class & Full & Training set & Test set\\\hline
    COVID-19
    & 3,616 & 2,893 & 723 \\
    Lung Opacity
    & 6,012 & 4,810 & 1,202 \\
    Normal
    & 10,192 & 8,154 & 2,038 \\
    Viral Pneumonia
    & 1,345 & 1,076 & 269 \\
    \hline
    \end{tabular}
\end{table}
\begin{table}[t]
    \centering
    \caption{Details of the COVID5K dataset~\cite{minaee2020deep}.}
    \label{tab2}
    \begin{tabular}{lccc}
    \hline
    Class & Full & Training set & Test set\\\hline
    COVID-19
    & 520 & 420 & 100 \\
    Normal
    & 5,000 & 2,000 & 3,000 \\
    \hline
    \end{tabular}
\end{table}
The datasets used in our study include the large COVID-19 CXR dataset~\cite{rahman2021exploring} and COVID5K dataset~\cite{minaee2020deep}, with COVID5K being an unbalanced dataset.
Table~\ref{tab1} shows that the COVID-19 CXR dataset has four categories, and the ratio of data in the training and test sets is 8:2.\footnote{https://www.kaggle.com/datasets/tawsifurrahman/covid19-radiography-database}
As presented in Table~\ref{tab2}, the unbalanced dataset has two classes with a total number of 5,520 images, where only 520 are COVID-19 images.\footnote{https://github.com/shervinmin/DeepCovid}
Figures~\ref{fig2} and~\ref{fig3} show some examples of CXR images from the two datasets.
Each image is grayscale and resized to a resolution of 224 pixels by 224 pixels.
The areas under the receiver operating characteristic curve (AUC), sensitivity (Sen), specificity (Spe), harmonic mean (HM) of Sen and Spe, and classification accuracy (Acc) were used as evaluation metrics.
COVID-19 was considered positive for Sen, Spe, HM, and AUC, whereas the others were considered negative.
\par
We used ResNet18 or ResNet50~\cite{he2016deep} as an encoder and stochastic gradient descent as an optimizer.
The projectors and predictor are two-layer MLPs with the same structure as that used in~\cite{grill2020bootstrap}.
After 40 epochs of self-supervised learning, 30 epochs of fine-tuning were performed on the datasets.
The results are the average and variance of the last 10 fine-tuning epochs.
\textcolor{black}{
In the self-supervised learning-based pretraining phase, the batch size, the generated view size, and the degree of moving average $\zeta$ were set to 256, 112, and 0.996, respectively~\cite{li2021self}.
Data augmentation methods such as cropping, resizing, flipping, and Gaussian blurring are used to generate random views.
In the batch knowledge ensembling-based fine-tuning phase, the hyperparameters $\omega$, $N$, $\lambda$, and $\tau$ were set to 0.5, 128, 8.0, and 1.0, respectively~\cite{li2022self}.
}
\par
\begin{table*}[t]
    \centering
    \caption{Test accuracy on the large COVID-19 CXR Dataset.}
    \label{tab3}
    \begin{tabular}{l|c|ccccc}
    \hline
    Method & Structure & Sen & Spe & HM & AUC & Acc \\\hline
    Ours & \multirow{10}*{ResNet50}
    & \bfseries{0.989$\pm$0.000} & \bfseries{1.000$\pm$0.000} & \bfseries{0.994$\pm$0.000} & \bfseries{1.000$\pm$0.000} & \bfseries{0.966$\pm$0.000}\\
    BKE &
    & 0.980$\pm$0.004 & 0.997$\pm$0.001 & 0.988$\pm$0.002 & 0.999$\pm$0.000 & 0.957$\pm$0.001 \\
    Cross &
    & 0.972$\pm$0.003 & 0.997$\pm$0.001 & 0.985$\pm$0.001 & 0.999$\pm$0.000 & 0.953$\pm$0.001 \\
    BYOL &
    & 0.973$\pm$0.004 & 0.996$\pm$0.001 & 0.985$\pm$0.002 & 0.999$\pm$0.000 & 0.954$\pm$0.001 \\
    SimSiam &
    & 0.974$\pm$0.004 & 0.995$\pm$0.001 & 0.984$\pm$0.002 & 0.998$\pm$0.000 & 0.950$\pm$0.001 \\
    PIRL-Jigsaw &
    & 0.977$\pm$0.003 & 0.997$\pm$0.001 & 0.987$\pm$0.001 & 0.999$\pm$0.000 & 0.951$\pm$0.001 \\
    PIRL-Rotation &
    & 0.973$\pm$0.002 & 0.997$\pm$0.001 & 0.985$\pm$0.001 & 0.999$\pm$0.000 & 0.951$\pm$0.001 \\
    SimCLR &
    & 0.913$\pm$0.006  & 0.994$\pm$0.001 & 0.952$\pm$0.003 & 0.996$\pm$0.000 & 0.936$\pm$0.001 \\
    Transfer &
    & 0.944$\pm$0.004 & 0.994$\pm$0.001 & 0.968$\pm$0.002 & 0.997$\pm$0.000 & 0.936$\pm$0.001 \\
    From Scratch &
    & 0.665$\pm$0.013 & 0.954$\pm$0.003 & 0.783$\pm$0.008 & 0.935$\pm$0.001 & 0.774$\pm$0.002 \\\hline
    Ours & \multirow{10}*{ResNet18}
    & \bfseries{0.982$\pm$0.000} & \bfseries{1.000$\pm$0.000} & \bfseries{0.994$\pm$0.000} & \bfseries{1.000$\pm$0.000} & \bfseries{0.960$\pm$0.001}\\
    BKE &
    & 0.972$\pm$0.004 & 0.998$\pm$0.000 & 0.985$\pm$0.002 & 1.000$\pm$0.000 & 0.951$\pm$0.001 \\
    Cross &
    & 0.944$\pm$0.003 & 0.990$\pm$0.001 & 0.967$\pm$0.001 & 0.996$\pm$0.000 & 0.934$\pm$0.002 \\
    BYOL &
    & 0.934$\pm$0.007 & 0.990$\pm$0.002 & 0.961$\pm$0.003 & 0.995$\pm$0.000 & 0.932$\pm$0.001 \\
    SimSiam &
    & 0.940$\pm$0.002 & 0.988$\pm$0.001 & 0.963$\pm$0.001 & 0.996$\pm$0.000 & 0.929$\pm$0.001 \\
    PIRL-Jigsaw &
    & 0.931$\pm$0.004 & 0.992$\pm$0.001 & 0.961$\pm$0.002 & 0.997$\pm$0.000 & 0.930$\pm$0.001 \\
    PIRL-Rotation &
    & 0.936$\pm$0.007 & 0.994$\pm$0.001 & 0.964$\pm$0.003 & 0.997$\pm$0.000 & 0.930$\pm$0.001 \\
    SimCLR &
    & 0.806$\pm$0.012  & 0.982$\pm$0.001 & 0.886$\pm$0.007 & 0.978$\pm$0.000 & 0.903$\pm$0.002 \\
    Transfer &
    & 0.900$\pm$0.008 & 0.981$\pm$0.003 & 0.939$\pm$0.003 & 0.993$\pm$0.000 & 0.909$\pm$0.001 \\
    From Scratch &
    & 0.849$\pm$0.010 & 0.958$\pm$0.004 & 0.900$\pm$0.004 & 0.974$\pm$0.000 & 0.831$\pm$0.001 \\\hline
    \end{tabular}
\end{table*}
\begin{figure*}[t]
        \centering
        \subfigure[]{
        \centering
        \includegraphics[width=6.5cm]{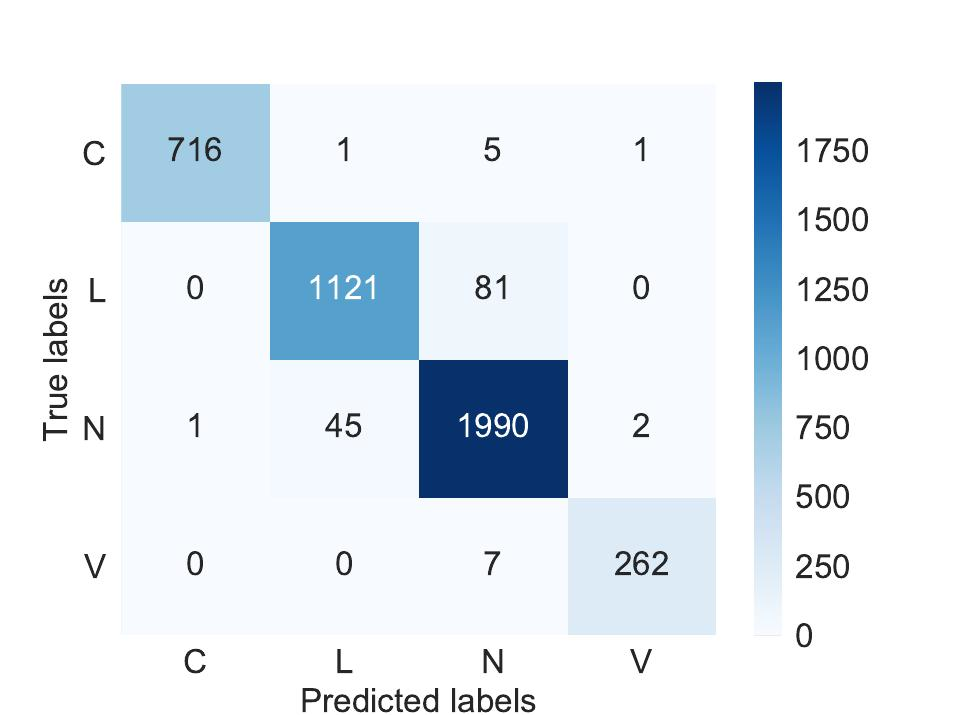}
        }
        \subfigure[]{
        \centering
        \includegraphics[width=6.5cm]{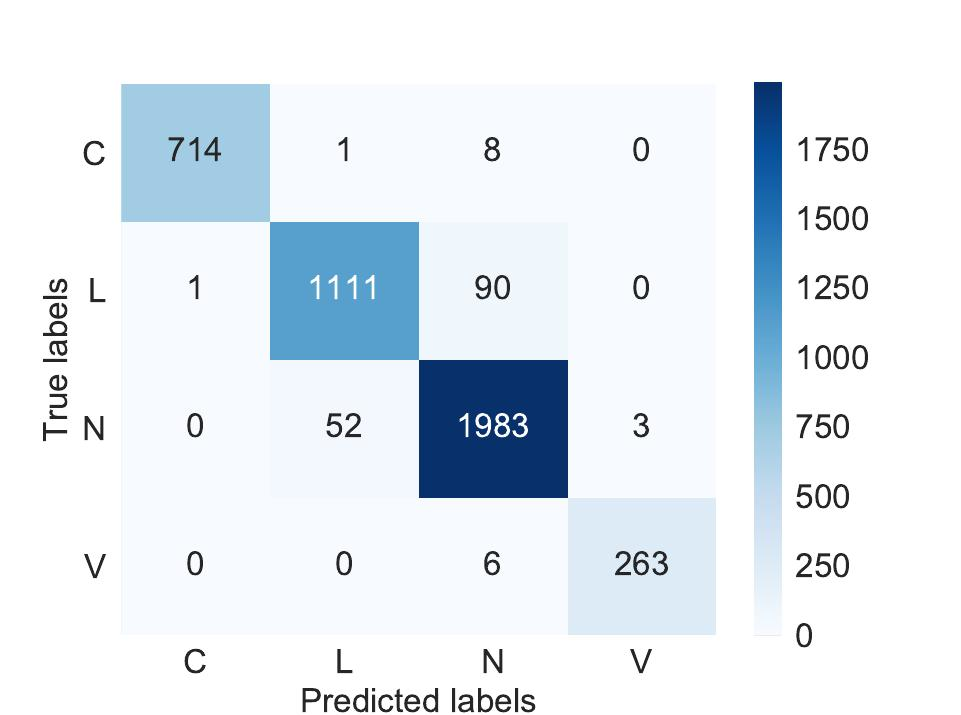}
        }
        \caption{Best performance confusion matrix of our method. (a): ResNet50, (b): ResNet18.}
        \label{fig4}
\end{figure*}
\begin{table*}[t]
    \centering
    \caption{Test accuracy in different annotated data volumes when compared with vision transformer-based methods.}
    \label{tab4}
    \setlength{\tabcolsep}{5.5mm}{
    \begin{tabular}{l|c|cccccc}
    \hline
    Method & Structure & 1\% & 10\% & 50\% & 100\% \\\hline
    Ours & ResNet50 
    & \bfseries{0.859} & \bfseries{0.934} & \bfseries{0.960} & \bfseries{0.966} \\
    Ours & ResNet18 
    & 0.811 & 0.925 & 0.952 & 0.960 \\
    RGMIM & ViT-Base
    & 0.771 & 0.919 & 0.957 & 0.962 \\
    MAE & ViT-Base
    & 0.754 & 0.903 & 0.948 & 0.956 \\
    Transfer & ViT-Base
    & 0.689 & 0.893 & 0.940 & 0.953 \\
    From Scratch & ViT-Base
    & 0.413 & 0.645 & 0.810 & 0.848 \\
    \hline
    \end{tabular}}
\end{table*}
\begin{figure*}[t]
        \centering
        \subfigure[]{
        \centering
        \includegraphics[width=6.5cm]{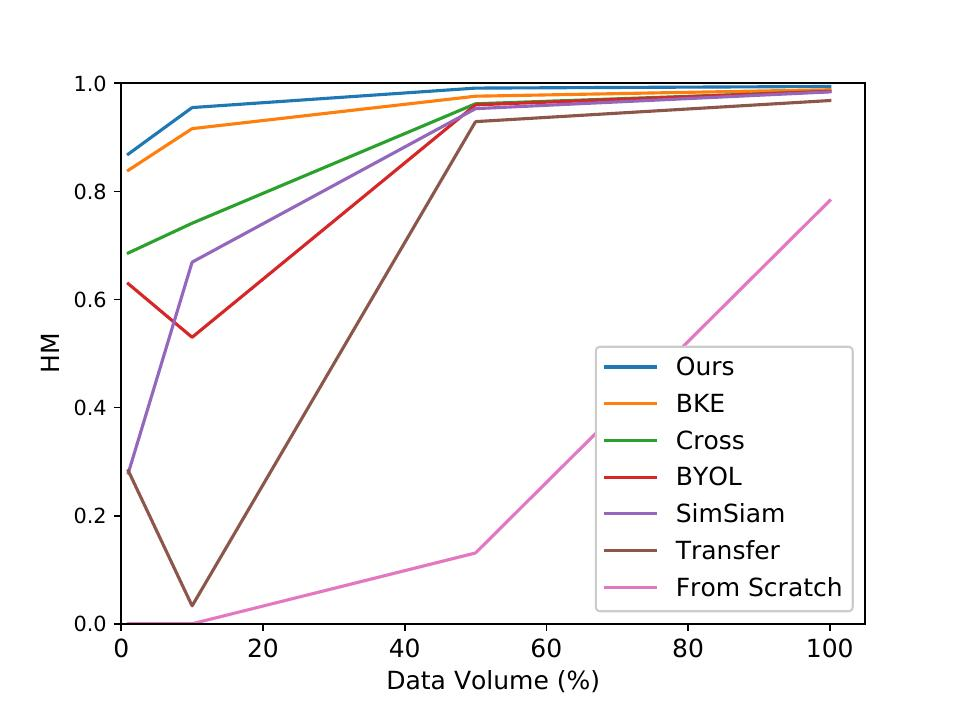}
        }
        \subfigure[]{
        \centering
        \includegraphics[width=6.5cm]{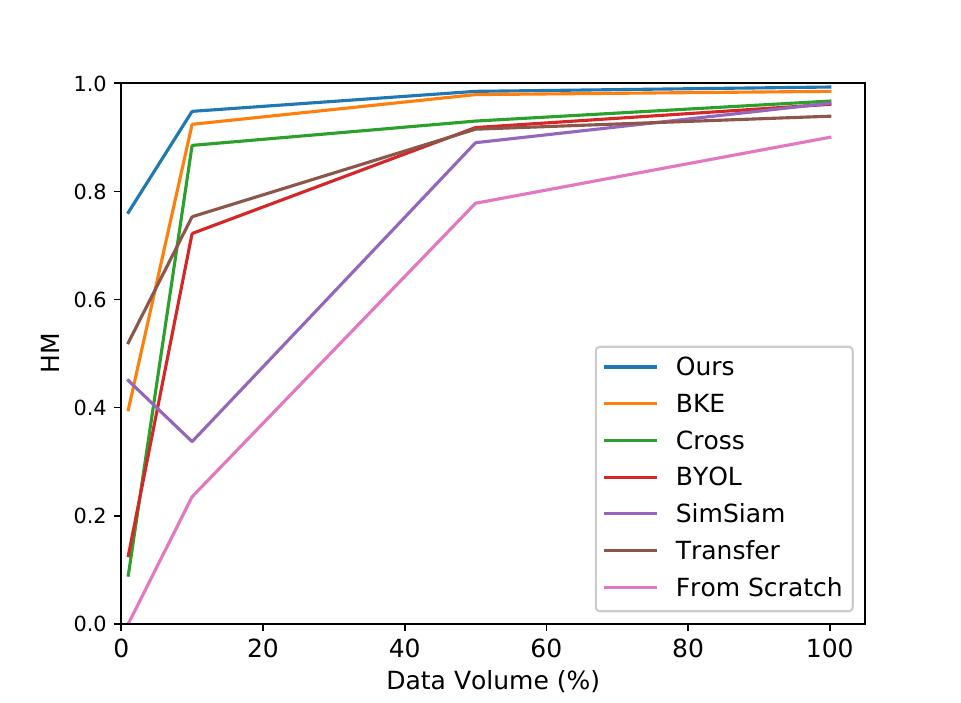}
        }
        \subfigure[]{
        \centering
        \includegraphics[width=6.5cm]{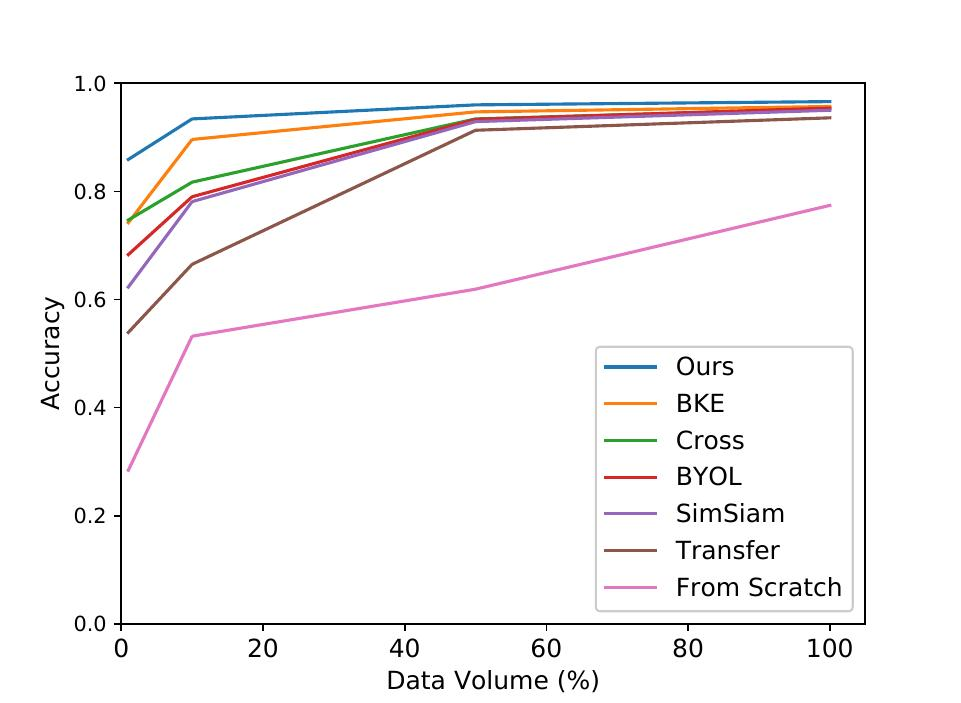}
        }
        \subfigure[]{
        \centering
        \includegraphics[width=6.5cm]{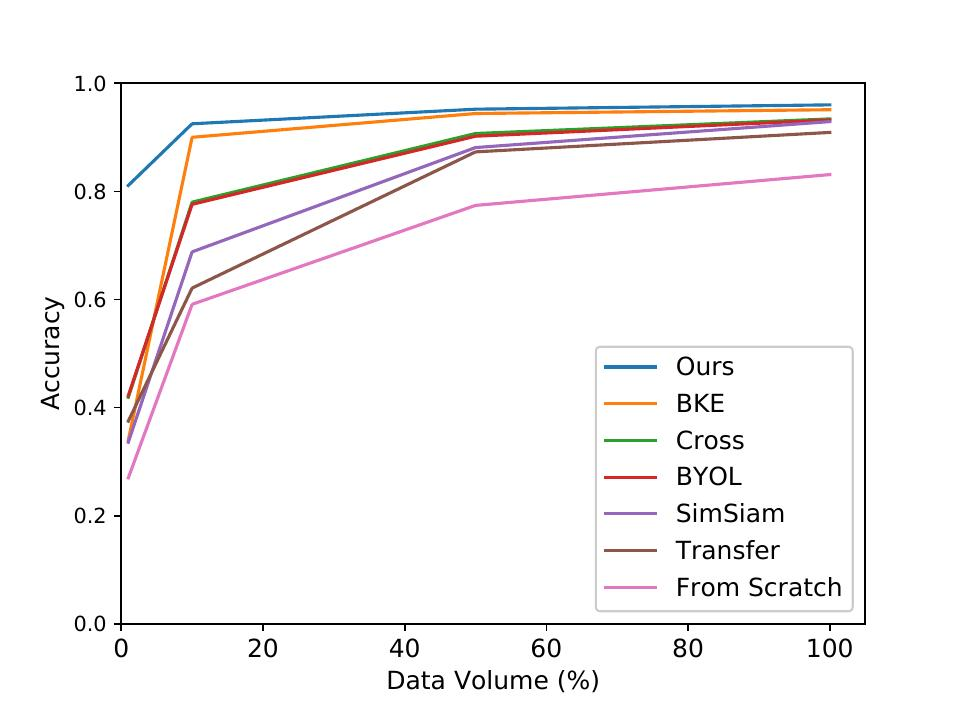}
        }
        \caption{Test accuracy in different annotated data volumes: (a) HM of ResNet50, (b) HM of ResNet18, (c) Accuracy of ResNet50, and (d) Accuracy of ResNet18.}
        \label{fig5}
\end{figure*}
\begin{table*}[t]
    \centering
    \caption{Test accuracy on the COVID5K dataset.}
    \label{tab5}
    \begin{tabular}{l|c|ccccc}
    \hline
    Method & Structure & Sen & Spe & HM & AUC \\\hline
    Ours &
    & 0.990$\pm$0.000 & 0.971$\pm$0.004 & \bfseries{0.980$\pm$0.002} & \bfseries{0.997$\pm$0.000} \\
    BKE &
    & 0.926$\pm$0.013 & \bfseries{0.989$\pm$0.001} & 0.957$\pm$0.007 & 0.995$\pm$0.000 \\
    Cross & ResNet50
    & \bfseries{0.999$\pm$0.005} & 0.925$\pm$0.016 & 0.960$\pm$0.008 & 0.995$\pm$0.000 \\
    Transfer &
    & 0.961$\pm$0.010 & 0.908$\pm$0.013 & 0.934$\pm$0.005 & 0.984$\pm$0.001 \\
    From Scratch &
    & 0.818$\pm$0.026 & 0.916$\pm$0.016 & 0.864$\pm$0.009 & 0.930$\pm$0.002 \\\hline
    Ours &
    & 0.958$\pm$0.004 & \bfseries{0.988$\pm$0.002} & \bfseries{0.973$\pm$0.002} & \bfseries{0.989$\pm$0.000} \\
    BKE &
    & 0.939$\pm$0.003 & 0.973$\pm$0.002 & 0.955$\pm$0.002 & \bfseries{0.989$\pm$0.000} \\
    Cross & ResNet18
    & \bfseries{0.970$\pm$0.000} & 0.946$\pm$0.007 & 0.958$\pm$0.004 & 0.987$\pm$0.000 \\
    Transfer &
    & 0.910$\pm$0.016 & 0.987$\pm$0.002 & 0.947$\pm$0.008 & 0.976$\pm$0.000 \\
    From Scratch &
    & 0.895$\pm$0.007 & 0.978$\pm$0.002 & 0.935$\pm$0.003 & 0.956$\pm$0.001 \\\hline
    \end{tabular}
\end{table*}
\begin{figure*}[t]
        \centering
        \subfigure[]{
        \centering
        \includegraphics[width=6.5cm]{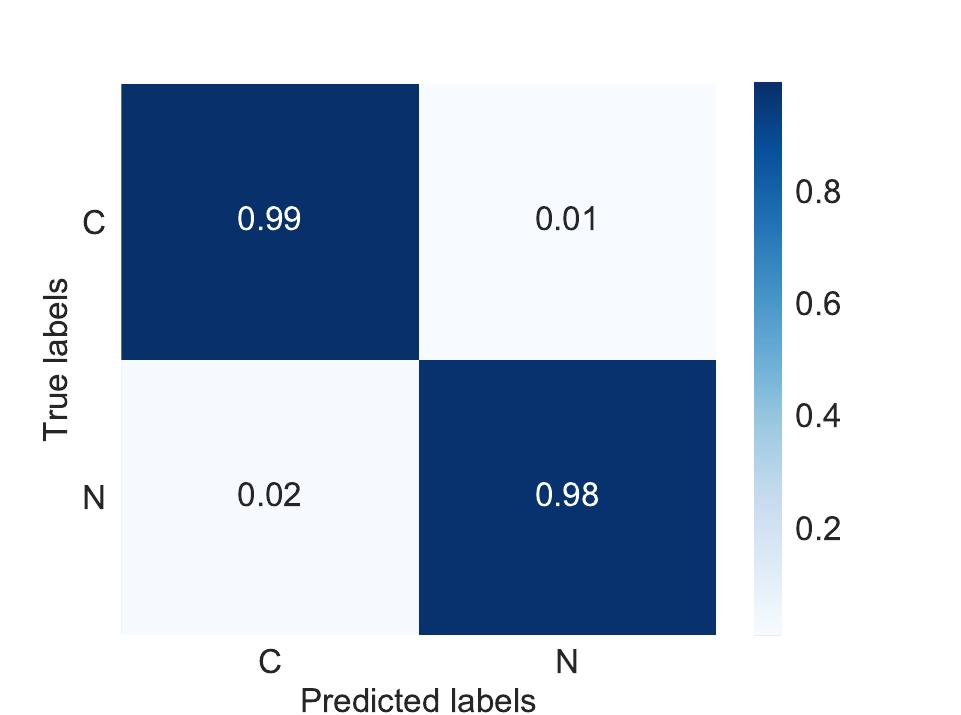}
        }
        \subfigure[]{
        \centering
        \includegraphics[width=6.5cm]{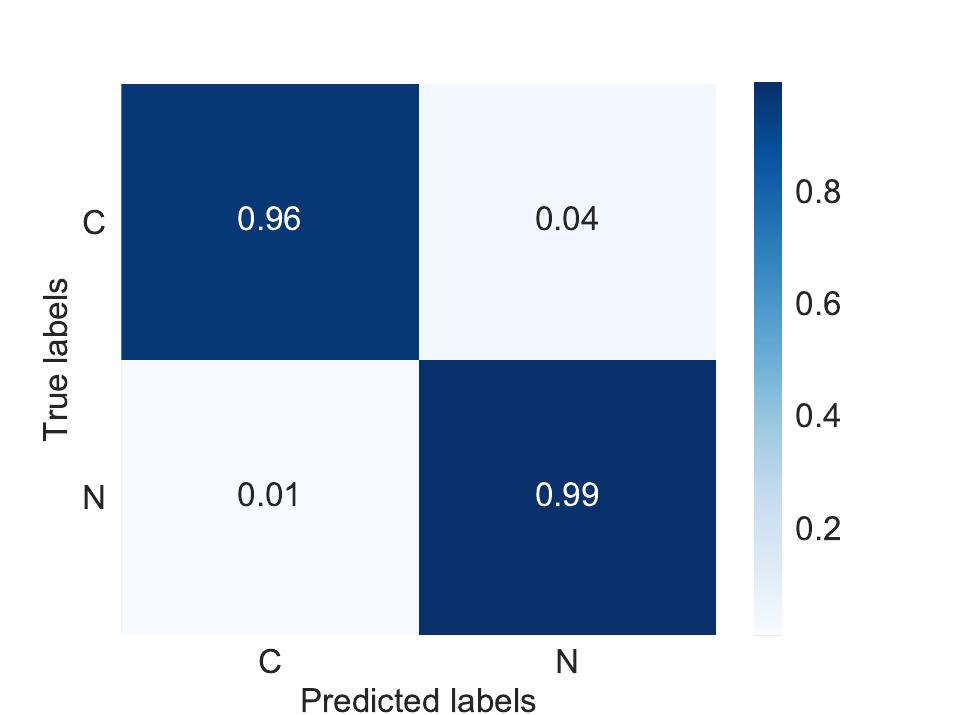}
        }
        \caption{Best performance confusion matrix of our method. (a): ResNet50, (b): ResNet18.}
        \label{fig6}
\end{figure*}
Several contrastive-based self-supervised learning methods were used, including BKE~\cite{li2022self}, Cross~\cite{li2022covid}, BYOL~\cite{grill2020bootstrap}, SimSiam~\cite{chen2021exploring}, PIRL~\cite{misra2020self}, and SimCLR~\cite{chen2020simple}.
\textcolor{black}{
Note that Cross~\cite{li2022covid} is our previous work that only considered self-supervised learning for COVID-19 detection.
}
As masked image modeling-based self-supervised learning methods using vision transformer~\cite{dosovitskiy2020vit} have recently become a new trend, we also use two SOTA methods RGMIM~\cite{li2022rgmim} and MAE~\cite{he2022masked} for comparison.
In our experiments, RGMIM and MAE used the ViT-Base model.
In addition, we trained from scratch and used transfer learning methods as baselines.
To test COVID-19 detection accuracy using a small amount of annotated data, we selected 1\%, 10\%, and 50\% of the training set for the fine-tuning process, respectively.
Note that we use the same selection ratio in each category.
\subsection{Test Accuracy on the Large COVID-19 CXR Dataset}
The test accuracy of COVID-19 detection on the training data is presented in Table~\ref{tab3}.
Specifically, when applying the ResNet50 model to all training data, transfer learning achieved HM, AUC, and Acc scores of 0.968, 0.997, and 0.936, respectively, and the best comparison method BKE~\cite{li2022self} achieved HM, AUC, and Acc scores of 0.988, 0.999, and 0.957, respectively.
On the other hand, our method achieved HM, AUC, and Acc scores of 0.994, 1.000, and 0.966, respectively, on the large COVID-19 CXR dataset.
\par
Figure~\ref{fig4} depicts the best performance confusion matrix of our method.
Our method not only discriminates well between patients with COVID-19 and normal patients but also achieves very high accuracy in identifying COVID-19 and other pneumonia.
The test results in different settings show that our method achieves promising detection results on the large COVID-19 CXR dataset, and our method significantly outperforms other comparison methods, improving COVID-19 detection performance.
\textcolor{black}{
Different from the previous method BKE~\cite{li2022self}, we introduce batch knowledge ensembling into the fine-tuning phase, reducing the computation cost and required memory.
Therefore, our method can run on a single NVIDIA Tesla P100 GPU with a memory of 16G, whereas the previous method necessitates two GPUs.
In addition, the training times of our method and BKE are approximately 97 and 124 min, respectively.
Our method is faster and more efficient than the previous method.
}
\par
The COVID-19 detection results for different annotated data volumes are presented in Table~\ref{tab4} and Fig.~\ref{fig5}.
The table and figure show that, compared with other comparison methods, our method vastly improved COVID-19 detection in a small amount of annotated data situations, such as 1\% and 10\% of the training set (169 and 1,693 images), and achieved promising detection performance even for 10\% of the training set.
Compared with the vision transformer-based methods RGMIM~\cite{li2022rgmim} and MAE~\cite{he2022masked}, although we use the traditional and straightforward ResNet model, our method outperformed them, especially when the amount of annotated data was significantly reduced.
\textcolor{black}{
In the real world, COVID-19 may have limited annotated training data because of the varying infection status, medical resources, and data-sharing policies of different countries~\cite{alhalaseh2021allocation}.
However, the proposed method can still be applied to this case for high-performance automatic COVID-19 detection.
}
\subsection{Test Accuracy on the COVID5K Dataset}
The test accuracy of the COVID-19 detection on the training data and confusion matrices of our method are presented in Table~\ref{tab5} and Fig.~\ref{fig6}.
The results show the average and variance of the last 10 fine-tuning epochs.
Specifically, when using all training data and the ResNet50 structure, transfer learning achieved Sen, Spe, and HM scores of 0.961, 0.908, and 0.934, respectively, and the best comparison method Cross~\cite{li2022covid} achieved Sen, Spe, and HM scores of 0.999, 0.925, and 0.960, respectively.
On the other hand, our method achieved Sen, Spe, and HM scores of 0.990, 0.971, and 0.980, respectively, on the unbalanced COVID-19 CXR dataset.
\textcolor{black}{
Our method achieved promising detection results on the unbalanced COVID-19 CXR dataset, which shows its robustness and that it can be used in extreme data situations in the real world.
}
\begin{table}[t]
    \small
    \centering
    \caption{Evaluation results on the changes of the ensembling weight $\omega$ and the batch size $N$.}
    \label{tab6}
    \begin{tabular}{lcc}
    \hline
    $\omega$ & HM & Acc  \\\hline
    0.1 & 0.996 & 0.964  \\
    0.3 & 0.996 & 0.967  \\
    0.5 & 0.994 & 0.966  \\
    0.7 & 0.994 & 0.967  \\
    0.9 & 0.993 & 0.967  \\
    \hline
    \end{tabular}
    \,\,\,\,\,\,
    \begin{tabular}{lcc}
    \hline
    $N$ & HM & Acc  \\\hline
    32 & 0.993 & 0.961  \\
    64 & 0.994 & 0.963  \\
    128 & 0.994 & 0.966  \\
    256 & 0.993 & 0.962  \\
    512 & 0.992 & 0.964  \\
    \hline
    \end{tabular}
\end{table}
\begin{table}[t]
    \small
    \centering
    \caption{Evaluation results on the changes of the temperature $\tau$ and the weighting factor $\lambda$.}
    \label{tab7}
    \begin{tabular}{lcc}
    \hline
    $\tau$ & HM & Acc  \\\hline
    2 & 0.994 & 0.964  \\
    4 & 0.994 & 0.965  \\
    8 & 0.994 & 0.966  \\
    16 & 0.994 & 0.964  \\
    \hline
    \end{tabular}
    \,\,\,\,\,\,
    \begin{tabular}{lcc}
    \hline
    $\lambda$ & HM & Acc  \\\hline
    0.5 & 0.994 & 0.963  \\
    1 & 0.994 & 0.966  \\
    2 & 0.994 & 0.963  \\
    4 & 0.992 & 0.960  \\
    \hline
    \end{tabular}
\end{table}
\subsection{Exploring the Impact of Hyperparameters on Experimental Results}
The evaluation results of different hyperparameters in the batch knowledge ensembling-based fine-tuning phase are shown in Tables~\ref{tab6} and~\ref{tab7}.
For ensembling, $\omega$ accounts for the propagation of knowledge between the anchor CXR image and other images within the same batch. 
As $\omega$ increases, the refined soft targets gain more information from other samples.
We studied the effects of the ensembling weight $\omega$ by changing it from 0.1 to 0.9. 
We observed that our method is insensitive to the ensembling weight.
Furthermore, when $\omega$ became larger, the HM scores decreased, but the accuracy increased, which indicated a greater bias toward the correct detection of normalcy and other pneumonia.
\par
Because our method uses the category information of different CXR images in a batch, we investigated the effects of the batch size by changing it from 32 to 512. 
As shown in the table, our method achieved the best results when the batch size was 128.
The predicted logits and soft targets are scaled using $\tau$.
By increasing $\tau$, the probability distribution over classes becomes smoother. 
A temperature change of $\tau$ from 2.0 to 16.0 was used to investigate the effect of the temperature value. 
Our method is insensitive to temperature and achieved the best results when $\tau$ was set to 8.0.
A $\lambda$ value is adopted to balance the cross-entropy loss and batch knowledge ensembling loss, which is generally set to 1.0.
We investigate the effects of the $\lambda$ value by varying it from 0.5 to 4.0. 
As shown in the table, our method achieved the best results when $\lambda$ was set to 1.0.
\par
\textcolor{black}{
In the real world, because there are differences in shooting equipment and patients in different regions or countries, if the model is very sensitive to changes in hyperparameters, a lot of time and money will likely be wasted on hyperparameter adjustments~\cite{zimmerer2019unsupervised}.
However, from the evaluation results of different hyperparameters, the proposed method is insensitive to changes in hyperparameters, which shows its potential to be used in a real-world clinical situation.
}
\subsection{Performance Comparison with Existing Methods}
\textcolor{black}{
The performance comparison with existing methods for COVID-19 detection from CXR images is presented in Table~\ref{tab8}.
As shown in Table ~\ref{tab8}, although these methods have achieved relatively high detection accuracy~\cite{narin2021automatic, waheed2020covidgan, ozturk2020automated, zhang2020covid19xraynet, tougaccar2020covid, gianchandani2020rapid, wang2020covid, gour2022uncertainty}, these studies have typically been evaluated on small COVID-19 CXR image datasets with only two or three classes and may have limitations when used in real clinical situations.
However, our method was evaluated on a large COVID-19 CXR dataset with 4 classes and 3,616 COVID-19 images and achieved promising detection performance.
In addition, our method uses only the most used simple structure, ResNet50, which has advantages in terms of reliability and practicality.
}
\begin{table*}[t]
    \centering
    \caption{\textcolor{black}{Performance comparison with the existing methods.}}
    \label{tab8}
    \begin{tabular}{llll}
    \hline
    Method & Structure & Dataset & Accuracy \\\hline
    Narin et al.~\cite{narin2021automatic} & Inception-ResNetV2 & COVID-19: 50,  & Two-class: 0.980 \\
    &  & Normal: 50 & \\
    Waheed et al.~\cite{waheed2020covidgan} & Auxiliary Classifier Generative  & COVID-19: 403, & Two-class: 0.950 \\
    & Adversarial Network & Normal: 721 & \\
    Ozturk et al.~\cite{ozturk2020automated} & DarkCovidNet & COVID-19: 127, & Two-class: 0.981 \\
    &  & Normal: 500 & \\
    Zhang et al.~\cite{zhang2020covid19xraynet} & ResNet34 & COVID-19: 189, & Three-class: 0.911 \\
    &  & Normal: 235, & \\
    &  & Viral Pneumonia: 63 & \\
    Togacar et al.~\cite{tougaccar2020covid} & Stacked models: & COVID-19: 295, & Three-class: 0.993\\
    & MobileNetV2, SqueenzeNet, & Normal: 65, & \\
    & SVM & Viral Pneumonia: 98 & \\
    Gianchandani et al.~\cite{gianchandani2020rapid} & Ensemble models: & COVID-19: 423, & Three-class: 0.962 \\
    & VGG16, ResNet152, & Normal: 1,579, & \\
    & DenseNet201 & Viral Pneumonia: 1,485 & \\
    Wang et al.~\cite{wang2020covid} & COVID-Net  & COVID-19: 358, & Three-class: 0.933 \\
    &  & Normal: 8,066, & \\
    &  & Viral Pneumonia: 5,538 & \\
    Gour et al. et al.~\cite{gour2022uncertainty} & UA-ConvNet  & COVID-19: 219, & Three-class: 0.988 \\
    &  & Normal: 1,341, & \\
    &  & Viral Pneumonia: 1,345 & \\
    Ours & ResNet50 & COVID-19: 3,616, & Four-class: 0.966 \\
    &  & Normal: 10,192, & \\
    &  & Viral Pneumonia: 1,345, & \\
    &  & Lung Opacity: 6,012 & \\
    \hline
    \end{tabular}
\end{table*}
\section{Discussion}
\label{sec5}
Using DL for computer-aided detection can reduce the burden on healthcare systems~\cite{bhattacharya2021deep, feki2021federated, dou2021federated}.
On the large and unbalanced COVID-19 CXR datasets, our method exhibited good automatic COVID-19 detection performance.
\textcolor{black}{
Significantly, when using a small amount of annotated training data for fine-tuning, our method outperformed the other SOTA methods.
Because of the entirely different infection status, the number of medical resources, and data sharing policies of COVID-19 in other countries and cities, there is a likelihood of limited annotated training data~\cite{peiffer2020machine, latif2020leveraging}.
Nevertheless, the proposed method can still be applied to this case for high-performance COVID-19 detection, which makes our method stands out above the pool of studies on DL for COVID-19 detection.
}
\par
\textcolor{black}{
Our method also has limitations.
For example, because our method is contrastive-based self-supervised learning and is designed on CNNs, we have not yet figured out how to migrate it to new vision transformer-based structures~\cite{liu2021swin, mehta2022mobilevit, li2022efficientformer}, which will be one of our future studies. 
In our experiments, we want to explore the impact and robustness of the initial parameters on different fine-tuning stages and data volumes. 
Hence, we use the average of the last 10 fine-tuning epochs to test the performance. 
Also, since we evaluate model performance in different subsets of settings (i.e., 1\%, 10\%, and 50\%), it is expensive to perform N-fold cross-validation in all settings. 
However, N-fold cross-validation is a more common method, which we will consider in our future work.
For the medical AI field, since there is already a trend to shift from only using AI-driven models to the Internet of Medical Things (IoMT) enabled systems~\cite{ghubaish2020recent}, one of our future directions is to apply our algorithms to such systems, which can perform high-efficiency real-time COVID-19 detection.
Furthermore, the ethical and privacy issues related to medical data sharing have been the main challenges for computer-aided detection systems~\cite{dhar2023challenges}.
However, our related studies about medical data distillation~\cite{li2020soft, li2022compressed, li2022ddpp, li2023sharing} can improve the effectiveness and security of medical data sharing among different medical facilities, which fits well with the proposed method and is expected to be applied in clinical situations.
}
\section{Conclusion}
\label{sec6}
\textcolor{black}{
We have proposed a novel automatic COVID-19 detection method with self-supervised learning and batch knowledge ensembling using CXR images.
Self-supervised learning-based pretraining can learn distinguished representations from CXR images without manually annotated labels.
Furthermore, batch knowledge ensembling-based fine-tuning can utilize category knowledge of images in a batch according to their visual feature similarities to boost detection performance.
On two public COVID-19 CXR datasets, including a large dataset and an unbalanced dataset, our method exhibited promising COVID-19 detection performance.
}
\par
\textcolor{black}{
In the real world, COVID-19 may have limited annotated training data because of the varying infection status, medical resources, and data-sharing policies of different countries.
The proposed method can still be applied to this case for high-performance automatic COVID-19 detection.
Also, there are differences in shooting equipment and patients in different regions or countries, if the model is very sensitive to changes in hyperparameters, a lot of time and money will likely be wasted on hyperparameter adjustments.
The proposed method is insensitive to changes in hyperparameters, which shows its potential to be used in a real-world clinical situation.
Our method can reduce the workloads of healthcare providers and radiologists.
}
\par
\textcolor{black}{
Despite the promising experimental results, the proposed method should be tested on other COVID-19 CXR image datasets and different image modalities ($e.g.$, CT and Ultrasound) for any potential bias.
Furthermore, comparing with traditional diagnostic methods, such as PCR tests and clinical assessments will be one of our future works.
Also, exploring the utilization of the proposed method on transformer-based structures is a potential research direction.
}
\section*{Declaration of competing interest}
None declared.
\section*{Acknowledgments}
This study was partly supported by the MEXT Doctoral program for Data-Related InnoVation Expert Hokkaido University (D-DRIVE-HU) program, the Hokkaido University-Hitachi Collaborative Education and Research Support Program and AMED Grant Number JP21zf0127004. This study was conducted at the Data Science Computing System of Education and Research Center for Mathematical and Data Science, Hokkaido University.
\bibliographystyle{elsarticle-num}
\bibliography{CBM}
\end{document}